%% file: main.tex
\documentclass[a4paper]{tufte-handout}

\usepackage{lab_notes}
\usepackage{tuftefoot}
\usepackage{hyperref}
\hypersetup{
    pdffitwindow=false,            
    pdfstartview={Fit},            
    pdftitle={Lab notes 2014},     
    pdfauthor={Your Name},         
    pdfsubject={},                 
    pdfnewwindow=true,             
    colorlinks=true,               
    linkcolor=DarkScarletRed,      
    citecolor=DarkChameleon,       
    filecolor=DarkPlum,            
    urlcolor=DarkSkyBlue           
}

\graphicspath{{figs/}}

\title[Public Algorithms Survey]{The Public Algorithms Survey in Allegheny County}
\date{January 15, 2023}  

\usepackage{booktabs}

\begin{document}
\maketitle


\begin{dummystart}
\end{dummystart}
\begin{fullwidth}
\input{010_introduction}

\input{011_key_observations}
\input{020_summary}
\input{030_methodology}

\input{040_acknowledgement}
\end{fullwidth}









\end{document}

%% file: 010_introduction.tex
\section{Overview}
Algorithms are increasingly used in public domains to assist in human decision-making\TFfootnotemark\TFfootnotetext{\label{fn1}Karen Levy, Kyla E Chasalow, and Sarah Riley. Algorithms and decision-making in the public sector. Annual Review of Law and Social Science, 17:309-334, 2021.}. These algorithms influence government functions ranging from day-to-day management, such as waste collection, to high-stakes, such as pretrial risk assessment, child welfare, and 
predictive policing\TFfootnotemark\TFfootnotetext{\label{fn1}Robert Brauneis and Ellen P Goodman. Algorithmic transparency for the smart city. Yale JL \& Tech., 20:103, 2018.}.
There has been growing research and media attention on the use of algorithmic decision-making aids, as well as how the public perceives such use. For example, a survey of American adults conducted by the Pew Research Center found that despite the growing use of a variety of computer algorithms in both public and commercial sectors, people were often skeptical of those tools for reasons such as privacy violation or failure to capture the nuance of complex situations\TFfootnotemark\TFfootnotetext{\label{fn1}Aaron Smith. Public attitudes toward computer algorithms. Pew Research Center, November 2018.}. 
In contrast, a recent survey based on a sample of the Dutch adult population reported that people often saw algorithmic decision-making systems to be fairer than human decision-makers\TFfootnotemark\TFfootnotetext{\label{fn1}Natali Helberger, Theo Araujo, and Claes H de Vreese. Who is the fairest of them all? public attitudes and expectations regarding automated decision-making. Computer Law \& Security Review, 39:105456, 2020.}. 
These seemly inconsistent results might suggest that people's opinions on algorithmic decision-making systems may be multi-dimensional, which also depends on the context of how systems are used in practice. 

When algorithms are used as part of the decision-making in public sectors by local governments, community members are the subject of those decisions. How do local communities make sense of the use of algorithms in their local government context? For the first time, a survey was conducted to provide insight into how residents are aware of the use of public algorithms in their local communities, their thoughts about key issues, and their experiences of interacting with the government sectors involving algorithmic decision-making.

The goal of this survey is to assess people's knowledge about, experience with, and attitudes toward the use of algorithmic tools in public sectors. The survey focuses on Allegheny County, the most populous county in southwestern Pennsylvania. The survey was conducted by a Pitt/CMU collaborative team and the University Center for Social and Urban Research at the University of Pittsburgh (UCSUR) in April 2021. Nearly 1,500 people living around Allegheny County identified from the UCSUR Research Registry participated. The survey data were adjusted to make the sample representative of the demographic characteristics of the county.

The responses offer the broad perspectives of country residents as a whole, as well as a glimpse of how race, age, level of education, gender, income, and whether people live in the City of Pittsburgh or the suburbs, tend to have different views and experiences related to the use of public algorithms.

\section{About This Report}

This report is completed through a collaboration among researchers from the University of Pittsburgh, Carnegie Mellon University, the Institute for Cyber Law, Policy, and Security at the University of Pittsburgh (Pitt Cyber) and UCSUR. Questions on the analysis can be directed to Yu-Ru Lin and Beth Schwanke at: \url{yurulin@pitt.edu} \& \url{beth.schwanke@pitt.edu}.

%% file: 011_key_observations.tex
\section{Key Observations}

\paragraph{\bf Level of knowledge}
The survey found a substantial lack of awareness among residents regarding the existence of algorithms used by the City or County to help make public-sector decisions -- only 8\% said that they had heard of a public algorithm used by their local governments. While a number of algorithms were used by the City and County at the time of the survey\TFfootnotemark\TFfootnotetext{\label{fn1}These algorithmic tools include Smart Traffic Lights, Predictive Policing, Fire Prediction, Child Welfare Screening, Pretrial Risk Assessment, Rapid Rehousing Tool, Screening Tool to Assess Family Needs for Childhood Resources, and Opioid Overdose Risk Assessment. We note that the Opioid Overdose Risk Assessment is in development, but not currently being used.}, most of the high-stakes algorithms used across public sectors were not widely known. 

\paragraph{\bf Experience and perceptions}
Residents heard of governments' use of algorithmic tools mostly through news and educational settings. ``Direct experiences'' (e.g., they or their friends/relatives were affected by a public algorithm's decision-making) were also listed among the major sources through which people become aware of the use of public algorithms. The awareness was nearly three times higher among residents who had prior interaction with any of the local government sectors, compared with those who had not.

Despite only a small portion reporting awareness of the government's use of algorithms, more than half of the respondents (56\%) said they were aware of public algorithms' impact on them.

\paragraph{\bf Attitudes toward government use of algorithmic tools}
The majority of survey respondents expressed high or some levels of concern about government use of algorithms. Most concerning aspects include ``transparency'' (73\%), ``lack of public involvement'' (65\%), and ``3rd party data access'' (65\%). 
With respect to algorithmic tools used in different public sectors, people were more concerned about the use in Criminal Justice and Social Services than that in Transportation (71\% approval of the use in Transportation vs. only 37\% approval in both Criminal Justice and Social Services).

There were also significant demographic differences in the public's attitudes toward government use of algorithmic tools. Traditionally marginalized groups expressed concerns about the use of algorithms in public sectors more strongly.


\section{Implications of This Study}
It has been over six years since the City of Pittsburgh and Allegheny County began to build algorithmic decision-making into government services. However, the local public's awareness of any of such technology adoption remains strikingly low. At the same time, concerns about the government's use of algorithmic tools, particularly in terms of insufficient transparency, lack of public involvement, and issues involving third-party data access were raised among community members, especially those who are traditionally marginalized. As has been repeatedly shown in AI research, the burden of bias in algorithmic systems is often borne by already marginalized populations\TFfootnotemark\TFfootnotetext{\label{fn1}Virginia Eubanks. Automating inequality: How high-tech tools profile, police, and punish the poor. St. Martin's Press, 2018.}\TFfootnotemark\TFfootnotetext{\label{fn1}Jan C. Weyerer and Paul F. Langer. Garbage in, garbage out: The vicious cycle of ai-based discrimination in the public sector. In Proceedings of the 20th Annual International Conference on Digital Government Research, pages 509-511, 2019.}. 
This study suggests that a basic level of transparency and accountability should be provided to increase the community's trust in the algorithmic decision-making processes, as well as hopefully improving the processes and systems themselves. When new technology is introduced, populations that will be affected by the system should be properly engaged to ensure the system's accountability. Balancing the risks and benefits of algorithmic decision making is a great challenge as government agencies increasingly adopt algorithmic tools. As such, more human-centered design approaches, such as value-sensitive design, inclusive design, and participatory design, could be introduced in the system design / data collection / review processes to help understand and mitigate systems' potential risks and adversarial effects.

%% file: 020_summary.tex
\newpage
\section{Summary of Findings}

\textbf{1. The survey found a substantial lack of awareness among residents regarding the existence of algorithms used by the City or County to help make public-sector decisions. }
\begin{itemize}
\item[] Overall, only 8\% of the residents said that they had heard of a public algorithm used by the City or County, 25\% were unsure if they had heard of a public algorithm, and the rest stated they did not.  The City residents were twice as likely to report knowing about the use of public algorithms compared with those living in the suburbs. (Fig.~\ref{fig:Q1})
\input{figs/fig1}
\end{itemize}
\clearpage

\noindent{\bf 2. Most of the high-stakes algorithms used in the public sectors were not widely known.}
\begin{itemize}
\item[] The ``Smart Traffic Lights'' (52\%) was by far the most common known public algorithmic tool among the residents. The  ``Predictive Policing Tool'' was the second most known algorithm (29\%). The rest of the algorithms had much lower awareness level, and 34\% of respondents reported that they had not heard any of the algorithms listed in the survey. (Fig.~\ref{fig:Q4})
\input{figs/fig3}
\end{itemize}
\clearpage

\noindent{\bf 3. News, education, and direct experience are the major sources of awareness regarding the use of public algorithms. The awareness is higher among residents who had prior interaction with local government sectors, compared with those who had not.}
\begin{itemize}
\item[] Nearly half of the survey respondents (48\%) who had heard of public sector algorithms listed news as one of their sources of information. Educational settings (e.g. classrooms, and workshops) (26\%) and direct experience (e.g. they or their friends/relatives were affected by a public algorithm's decision making) (24\%) were listed as the second and third most common information sources. (Fig.~\ref{fig:Q1b})
\input{figs/fig2}
\item[] Respondents who reported having direct experience and interaction with public sectors were significantly more likely to be aware of the use of public algorithms than those who did not. People who reported interactions with a public sector were 2.8 times more likely to be aware of any algorithms, compared with people who had no prior interactions. Moreover, people who had interactions with a particular sector were 1.2--14.6 times more likely to be aware of algorithms relevant to that sector (from 1.2 in Transportation to 14.6 in the Public Housing sector), compared with people who had no prior interaction with that specific sector. (Fig.~\ref{fig:Q1_Q4_byQ3} and \ref{fig:Q4_byQ3}).
\input{figs/figA10}
\input{figs/figA11}
\end{itemize}
\clearpage

\noindent{\bf 4. More than half of the respondents said they were aware of public algorithms' impact on them.}
\begin{itemize}
\item[] Despite only a small portion (8\%) reporting awareness regarding the use of public algorithms, more than half of the survey respondents (56\%) expressed awareness of a high or some level of impact posed by public algorithms on their everyday life. (Fig. \ref{fig:Q2})
\input{figs/fig4}
\end{itemize}
\clearpage

\noindent{\bf 5. The majority of survey respondents expressed high or some levels of concern about the government's use of algorithms, especially on the ``transparency'' aspect.}
\begin{itemize}
\item[] The top three aspects where respondents expressed the greatest concerns are ``Transparency'' (little is made known about how an algorithm works, 73\%), ``Lack of public involvement'' (not be enough public involvement in creating algorithmic tools, 65\%), and ``3rd party data access'' (data from algorithmic tools could be shared with or accessed by others such as law enforcement or outside parties, 65\%). (Fig.~\ref{fig:Q5})\TFfootnotemark\TFfootnotetext{\label{fn3}The full description of each of the aspects in the survey:
\begin{itemize}
    \item Transparency: {\it I am concerned with ``black boxes'' where little is made known about how an algorithm works.}
    \item Lack of public involvement: {\it I am concerned that there will not be enough public involvement in creating and maintaining algorithmic tools.}     
    \item 3rd party data access: {\it I am concerned that data from algorithmic tools could be shared with or accessed by others (for example, law enforcement or outside parties).} 
    \item Data bias: {\it I am concerned with bias being present in the data or design of algorithmic tools.}
    \item Goverment use: {\it I am concerned with certain government agencies / sectors using algorithmic tools for any decision making.)}
    \item Loss of human control: {\it I am concerned that algorithmic tools increase loss of human control and judgement.}
\end{itemize} }
\input{figs/fig5}
\end{itemize}
\clearpage

\noindent{\bf 6. There was a drastic difference in the approval of public algorithms' use in the particular sector. People were more concerned about the use of algorithms in Criminal Justice and Social Services than that in Transportation.}
\begin{itemize}
\item[] 71\% of respondents agreed or strongly agreed on the public algorithms' use in Transportation, 55\% of respondents agreed or strongly agreed on that in Public Health, compared with a much lower level of approval for education, public housing, social services and criminal justice. (Fig.~\ref{fig:Q6}) 
\input{figs/fig6}
\end{itemize}
\clearpage

\noindent{\bf 7. Majority of residents said there was not sufficient accountability and transparency over governments' use of algorithmic tools.}
\begin{itemize}
\item[] Over half of the respondents disagreed that there is enough accountability and transparency about the use of algorithmic tools by both the City of Pittsburgh and Allegheny County, and more than 40\% of respondents considered oversight insufficient over both governments' use of algorithmic tools. (Fig.~\ref{fig:Q7} \& \ref{fig:Q8})
\input{figs/fig7}
\end{itemize}
\clearpage

\noindent{\bf 8. There were significant demographic differences in the public's attitudes toward governments' use of algorithmic tools. Traditionally marginalized groups expressed concerns about the use of algorithms in public sectors more strongly.}
\begin{itemize}
\item[] Marginalized groups, such as non-White respondents and gender minority groups, not only had a higher level of awareness of governments' use of algorithms, but also a greater level of disapproval for the use. For example, non-White respondents were 1.87 times more likely to be aware of the use of public algorithms, compared with the White respondents, and gender minorities were 4.38 times less likely to be unaware of any use of public algorithms. Across different public sectors, non-White respondents were 2-3 times more likely to strongly disagree that government should use algorithms in those sectors compared to White respondents. A much higher proportion of strong disagreement was observed in the non-White groups compared with White regarding Public Housing (11-15\% vs. 4\%), Social Services (11-13\% vs. 6\%), and Criminal Justice (17-21\% vs. 7\%). (Fig.~\ref{fig:Q6_eth})
\input{figs/fig8}
\end{itemize}
\clearpage

\clearpage

%% file: figs/fig1.tex

\begin{figure}
\paragraph{Do you know of any algorithms that the City of Pittsburgh and Allegheny County uses to make or help make decisions in the public sector?}
  \includegraphics{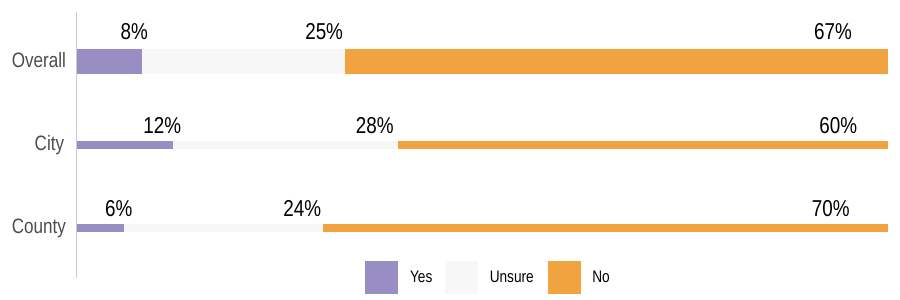}
  \caption{Awareness of the use of public algorithms.}
  \label{fig:Q1}
  \setfloatalignment{b}
\end{figure}

%% file: figs/fig3.tex
\begin{figure}
\paragraph{Below is a list of algorithms being developed, in-use, or previously in-use locally. Before taking this survey,  which public algorithms have you heard about in the City of Pittsburgh or Allegheny County? (Select all that apply)}
  \includegraphics{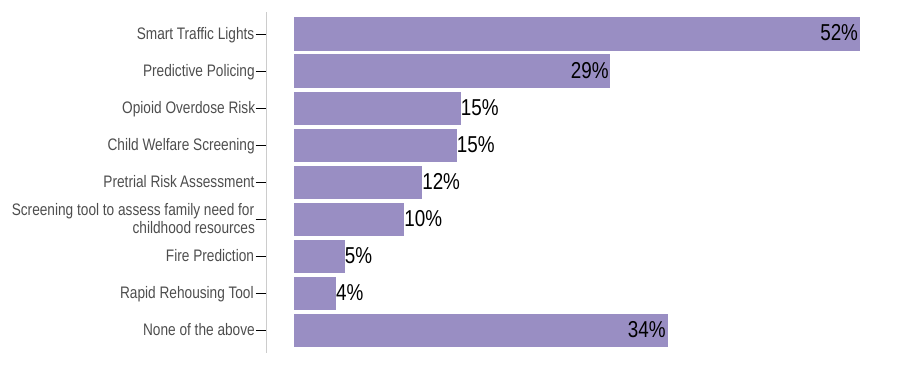}
  \caption{Awareness of algorithms in use locally.}
  \label{fig:Q4}
  \setfloatalignment{b}
\end{figure}

%% file: figs/fig2.tex
\begin{figure}
\paragraph{How did you become aware of these algorithms? (Select all that apply)}
  \includegraphics{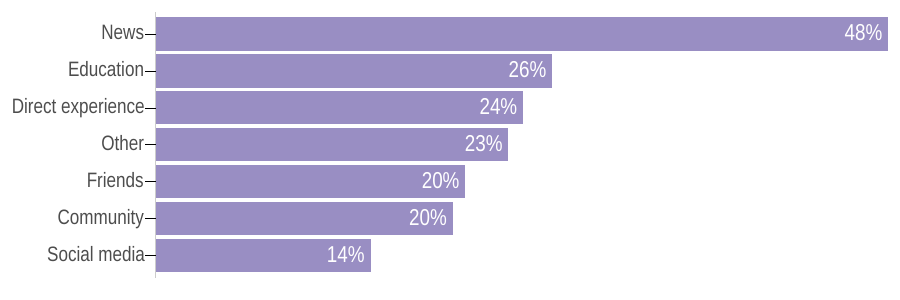}
  \caption{Source of awareness.}
  \label{fig:Q1b}
  \setfloatalignment{b}
\end{figure}

%% file: figs/figA10.tex
\begin{figure}
\paragraph{Do you interact with any of the following local government sectors? (Select all that apply)}
  \includegraphics{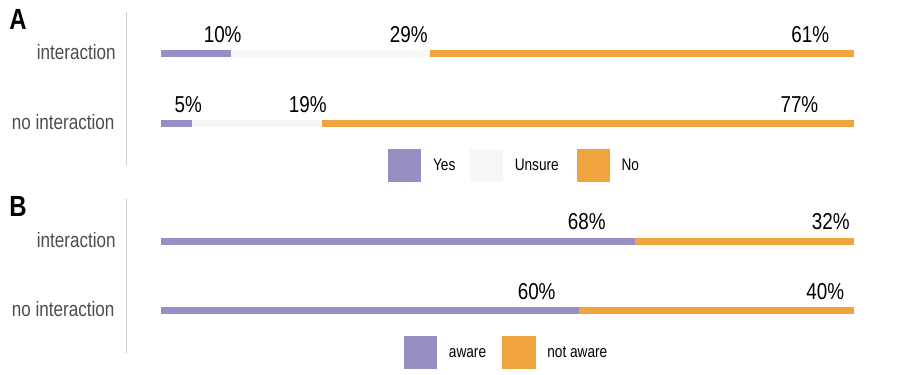}
  \vspace{1em}
  \caption{Awareness of (A) any public algorithms, and (B) specific algorithms in use locally, by whether or not one has any interactions with government sectors.}
  \label{fig:Q1_Q4_byQ3}
  \setfloatalignment{b}
\end{figure}

%% file: figs/figA11.tex
\begin{figure}
\paragraph{Do you interact with any of the following local government sectors? (Select all that apply)}
  \includegraphics{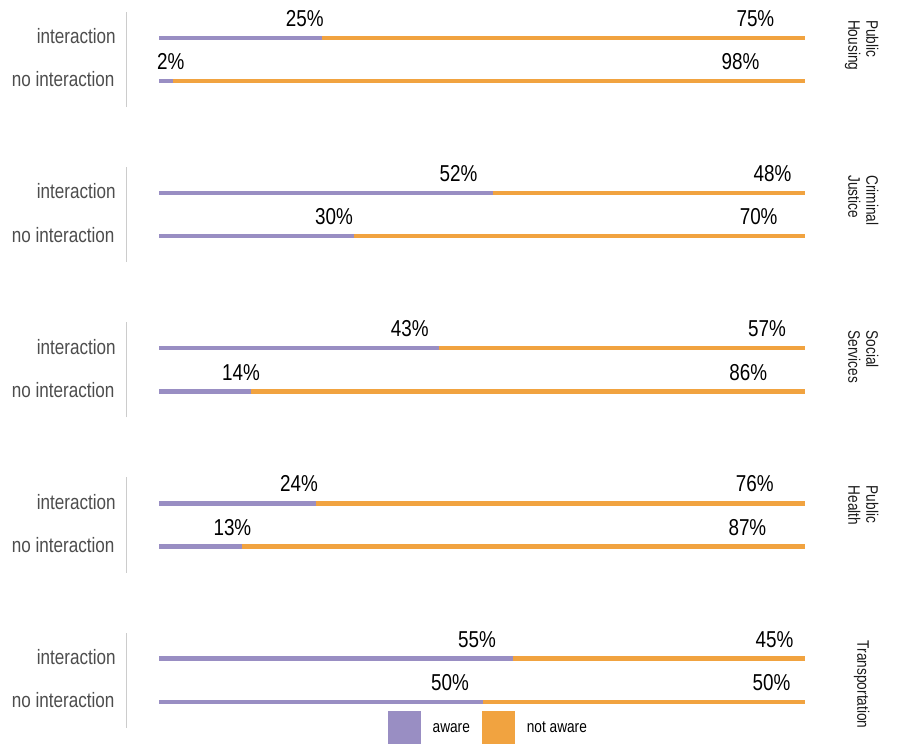}
  \vspace{1em}
  \caption{Awareness of algorithm in use by whether or not one has prior interaction with a particular sector.}
  \label{fig:Q4_byQ3}
  \setfloatalignment{b}
\end{figure}

%% file: figs/fig4.tex
\begin{figure}
\paragraph{What level of impact do public algorithms have on you (e.g. safety, health and finances) to your knowledge?}
  \includegraphics{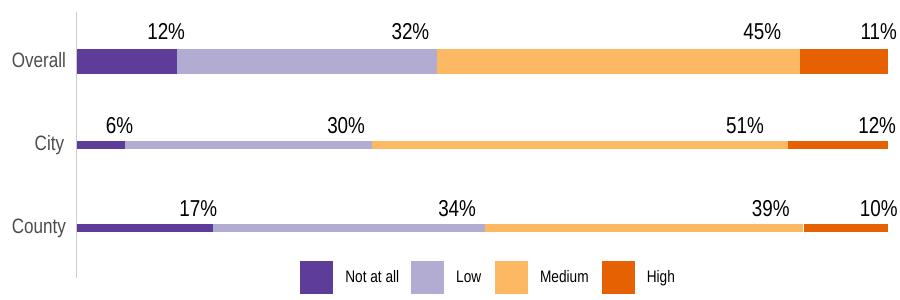}
  \caption{Perception of algorithms' impact.}
  \label{fig:Q2}
  \setfloatalignment{b}
\end{figure}

%% file: figs/fig5.tex
\begin{figure}
\paragraph{
How much do you agree with the following statements about the government's use of algorithmic decision-making tools? }
  \includegraphics{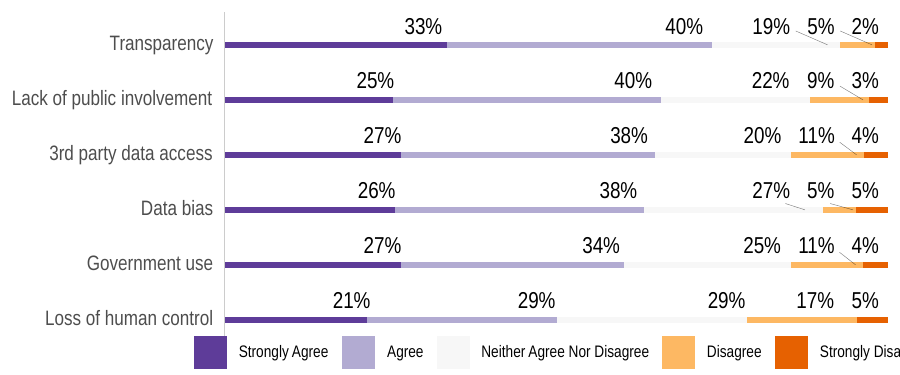}
  \vspace{1em}
  \caption{Concerns about the use of algorithmic decision-making tools.}
  \label{fig:Q5}
  \setfloatalignment{b}
\end{figure}

%% file: figs/fig6.tex
\begin{figure}
\paragraph{Governments should use algorithms in the following public sector(s):}
  \includegraphics{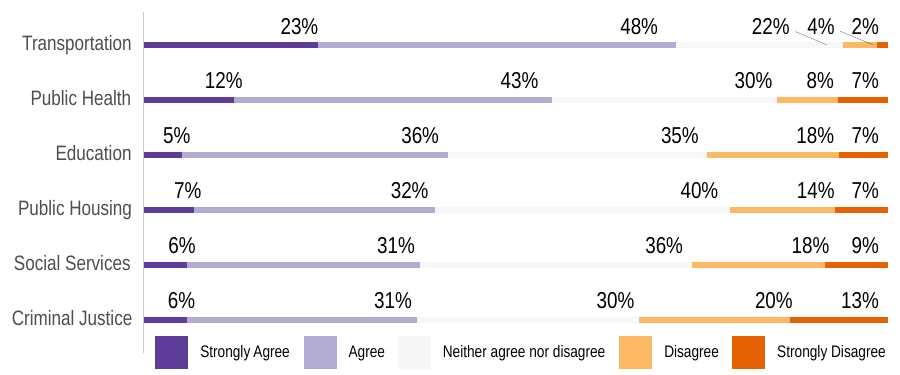}
  \vspace{1em}
  \caption{Agreement to use algorithms across different public sectors.}
  \label{fig:Q6}
  \setfloatalignment{b}
\end{figure}

%% file: figs/fig7.tex
\begin{figure}
\paragraph{How much do you agree with the following statements about the City of Pittsburgh's use of algorithmic tools?
There is enough \rule{1cm}{0.15mm} over government use of algorithmic tools.}
  \includegraphics{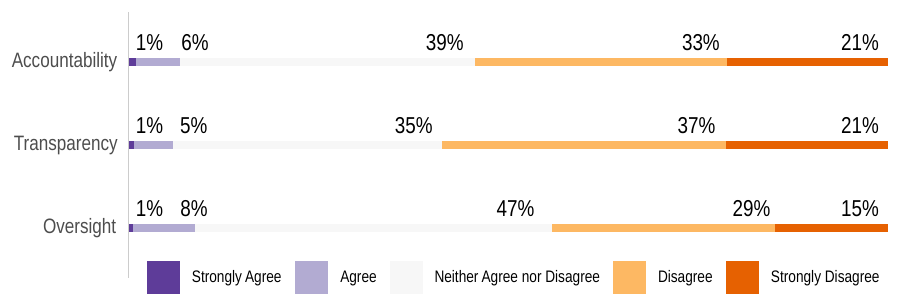}
  \vspace{1em}
  \caption{Attitude toward City's use of algorithms.}
  \label{fig:Q7}
  \setfloatalignment{b}
\end{figure}

\begin{figure}
\paragraph{How much do you agree with the following statements about Allegheny County's use of algorithmic tools?
There is enough \rule{1cm}{0.15mm} over government use of algorithmic tools.}
  \includegraphics{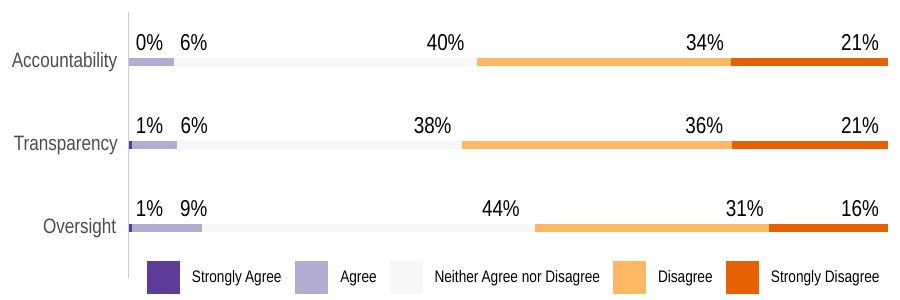}
  \vspace{1em}
  \caption{Attitude toward County's use of algorithms.}
  \label{fig:Q8}
  \setfloatalignment{b}
\end{figure}

%% file: figs/fig8.tex
\begin{figure}
\paragraph{Governments should use algorithms in the following public sector(s):}
  \includegraphics[width=1.6\linewidth]{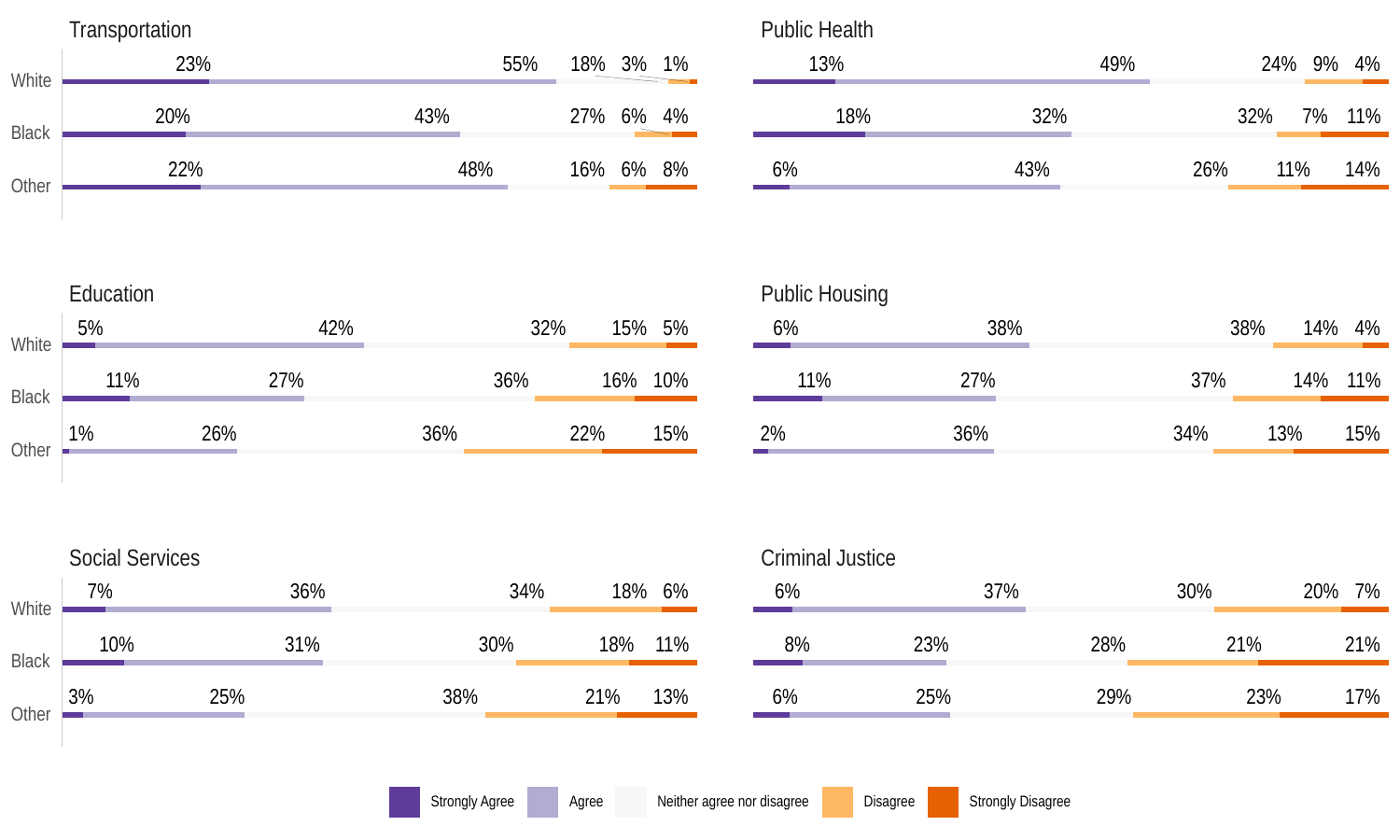}
  \vspace{1em}
  \caption{Ethnicity breakdown: Agreement to use algorithms across different public sectors.}
  \label{fig:Q6_eth}
  \setfloatalignment{b}
\end{figure}

%% file: 030_methodology.tex
\newpage
\section{Methodology}

An online survey was conducted between April 15 and May 4, 2021. This survey was sent to members of the UCSUR Research Registry\TFfootnotetext{\label{fn2}The full description of the registry is available from the UCSUR website
\href{https://ucsur.pitt.edu/registry.php}{UCSUR (@PittCSUR) -- PuRR Registry}} residing in Allegheny County for whom an email address was available. The registry contains approximately 10,000 community members recruited through UCSUR surveys and research studies who have agreed to be contacted for future research studies. Studies used to recruit registry members employed a combination of probability and nonprobability sampling designs. Email addresses are available for approximately 80\% of the registry members. Those without emails tend to be older, African American, and less educated. However, these demographic factors were adjusted for the weighting of the data (see below).

On April 15 an initial email invitation with a link to the survey was sent to 7,982 registry members from Allegheny County. The survey contained approximately 17 questions covering residents' awareness, experience, and attitudes toward the use of algorithmic tools by the local governments including the City of Pittsburgh and Allegheny County. The survey allows respondents to come back later to finish their responses. The median time respondents took to complete the survey is 11.9 minutes. Follow-up reminders to non-respondents were sent on April 21, 27, and 30, with a notice that data collection would end on May 4th. This resulted in 1,566 completed surveys, a 19.6\% response rate. 

\input{figs/figA0}

Note that this is technically a non-probability sampling method, as registry / panel members are volunteers. Thus, statements about ``margin of error'' are not appropriate, as these apply only to probability sampling designs. To reduce potential bias, the data were adjusted using statistical ``raking'' methods for age, gender, ethnicity, education, and income to make the sample more representative of the Allegheny County population aged 18 and older. The technique involves using the most recent American Community Survey's (ACS) age, gender, education, income, and ethnicity data for Allegheny County to compute case weights to ensure the sample mirrors the population distributions for each demographic variable. This is the standard approach in the survey industry when relying on survey registries and panels for population estimates.

%% file: figs/figA0.tex
\begin{figure}
  \includegraphics[width=1.4\linewidth]{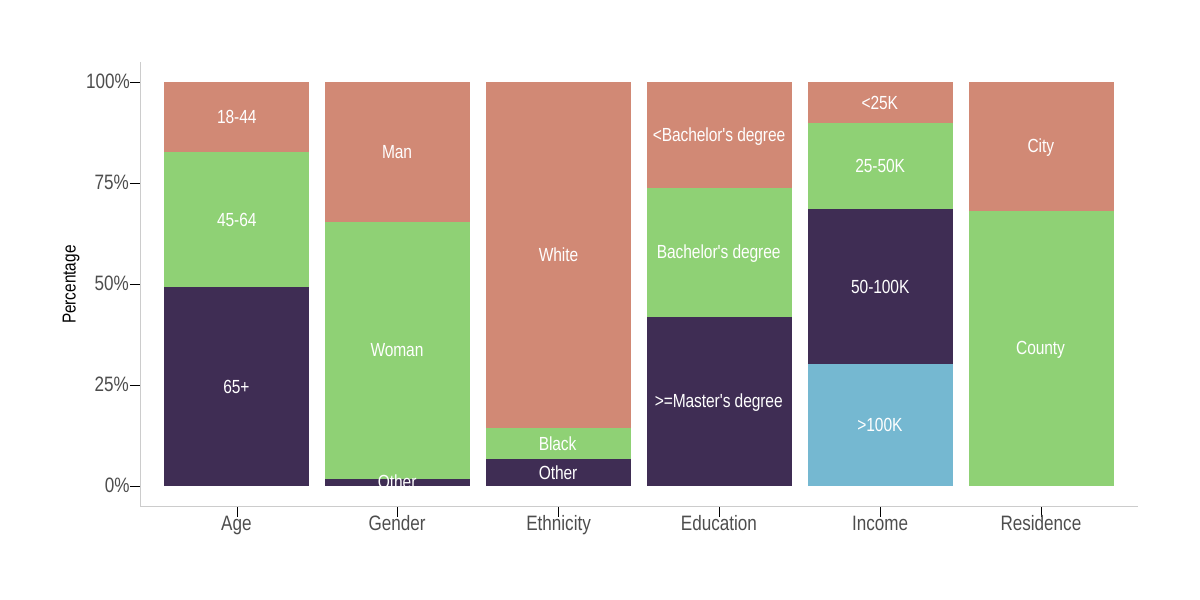}
  \caption{Respondents' demographics breakdown.}
  \label{fig:A0}
  \setfloatalignment{b}
\end{figure}

%% file: 040_acknowledgement.tex
\newpage
\section{Contributors}

\paragraph{University of Pittsburgh}
\begin{itemize}
\item Yu-Ru Lin, Associate Professor, School of Computing and Information
\item Beth Schwanke, Executive Director, Pitt Cyber

\item Rosta Farzan, Associate Professor, School of Computing and Information
\end{itemize}

\paragraph{Carnegie Mellon University}
\begin{itemize}
\item Bonnie Fan, former Ph.D. student, Carnegie Mellon University 
\item Motahhare Eslami, Assistant Professor, Human-Computer Interaction Institute
\item Hong Shen, Assistant Research Professor, Human-Computer Interaction Institute 
\item Sarah Fox, Assistant Professor, Human-Computer Interaction Institute
\end{itemize}

\paragraph{University of Pittsburgh, University Center for Social and Urban Research (UCSUR)}
\begin{itemize}
\item Scott R. Beach, Interim director \& director of survey research
\item Rhaven Nelson, Research database programmer
\end{itemize}